\UseRawInputEncoding
\documentclass{optica-article}

\journal{opticajournal} 

\articletype{Research Article}

\usepackage{lineno}

\begin{document}

\title{Selective Passive Tuning of Cavity Resonance by Mode Index Engineering of the Partial Length of a Cavity}

\author{Mohit Khurana,\authormark{1,2,*} Sahar Delfan ,\authormark{1,2} and Zhenhuan Yi\authormark{1,2}}

\address{\authormark{1} Department of Physics and Astronomy, Texas A\&M University, College Station, TX 77843, USA\\
\authormark{2}Institute of Quantum Science and Engineering, Texas A\&M University, College Station, TX  77843, USA\\}

\email{\authormark{*}mohitkhurana@tamu.edu} 


\begin{abstract*} 
Cavities in large-scale photonic integrated circuits often suffer from a wider distribution of resonance frequencies due to fabrication errors. It is crucial to adjust the resonances of cavities using post-processing methods to minimize the frequency distribution. We have developed a concept of passive tuning by manipulating the mode index of a portion of a microring cavity. Through analytical studies and numerical experiments, we have found that depositing a thin film of dielectric material on top of the cavity or etching the material enables us to fine-tune the resonances and minimize the frequency distribution. This versatile method allows for the selective tuning of each cavity's resonance in a large set of cavities in a single fabrication step, providing robust passive tuning in large-scale photonic integrated circuits. We show that proposed method achieves tuning resolution below 1/Q and range upto $10^{3}/Q$ for visible to near-infrared wavelengths. Furthermore, this method can be applied and explored in various optical cavities and material configurations.
\end{abstract*}

\section{Introduction}
On-chip microring resonators are widely used in various fields due to their small size, high-quality resonances, and advantages in precise manipulation of light \cite{Vahala2003}. In telecommunications, they enable wavelength division multiplexing (WDM) and modulation of light for increased data transmission over optical fibers \cite{Luo2014}. In data centers, they are used in optical interconnects for high-speed and energy-efficient communication \cite{Jrgensen2022, Shi2022}. They are also crucial in biosensing for label-free detection of biomolecules and in environmental monitoring to detect trace chemicals \cite{Bryan2023}. In quantum computing, they are used to generate entangled photon pairs and single-photon sources \cite{Engin2013, Ma2023, Savanier2016}, while in LIDAR systems, they improve autonomous vehicle navigation through frequency comb generation \cite{Lukashchuk2022, Chen2023}. Microring resonators are integral to on-chip spectrometers \cite{Zheng2019, Chen2023_2}, optical signal processing \cite{Stassen2019}, and biomedical imaging \cite{Chen2009}, particularly in Optical Coherence Tomography (OCT) \cite{Zhang2022}. They are also used in neuromorphic computing for AI \cite{Ohno2022}, atomic clock stabilization \cite{Zhou2024}, and the generation of supercontinuum light sources \cite{Suwanarat2018, Song2022}. In astrophotonics, they assist in exoplanet detection \cite{Suh2018}, and in optical computing and photonic neural networks, they form reliable fundamental elements of chip \cite{Zhang2024, Wu2022}. Moreover, microring resonators are used in terahertz generation \cite{Katti2018, Zhang2019} and secure photonic communication \cite{Chanana2022}. They also facilitate RF signal processing in microwave photonics, soil, and crop monitoring in agriculture \cite{Lo2017, CardenosaRubio2019}, and precision measurements in optical frequency metrology \cite{Moille2024}. Coupled-resonator optical waveguide (CROW) devices have advantages in development of complex photonic devices for non-linear photonics, optical band-pass filters, chip-integrated lasers, and sensors. These applications demonstrate the broad impact of microring resonators and PICs across industries, from healthcare and environmental monitoring to advanced computing and telecommunications.
\\

\noindent
In developing microring resonators, achieving the desired resonant frequencies is essential. However, deviations from the ideal design can occur due to various factors in the fabrication process. These fabrication errors, such as dimensional variations, sidewall roughness, material inhomogeneity, and coupling variability, significantly impact resonator performance by causing shifts and distributions in resonant frequencies \cite{Qin2013}. Dimensional variations in the radius, waveguide width, and thickness alter the optical path length and effective refractive index, leading to frequency shifts. Sidewall roughness from imperfect etching introduces scattering losses and refractive index fluctuations, while material inhomogeneity from non-uniform deposition or impurities further shifts the effective refractive index. The inconsistent microring and coupling waveguide gaps affect coupling efficiency, influencing the resonant frequency and resonance characteristics. To mitigate these issues, tight process control, such as high-precision lithography and etching processes, can minimize dimensional variations and roughness. Various post-fabrication tuning techniques, including thermal tuning \cite{Cunningham2010}, the electro-optic effect \cite{Guarino2007}, and localized material modification \cite{Larson2021, Bachman2011}, can adjust the refractive index and correct minor frequency shifts after fabrication \cite{Li2015}. Additional methods, such as controlled refractive index changes \cite{Carver2021}, stress application \cite{Chew2010}, selective material deposition or removal, localized heating, or mechanical deformation, provide powerful tools to optimize cavity performance. By compensating for fabrication inconsistencies, these post-fabrication techniques enhance device functionality and improve manufacturing yield without costly redesigns or rebuilds. Furthermore, post-processing methods offer a pathway to compensate for manufacturing variations and environmental effects, ensuring consistent performance across multiple devices. However, these methods are low-quality tuning, require live resonance monitoring, are hard to implement on multiple cavities at a time, and consume significant electric or optical power. Therefore, the problem of minimizing the resonance distribution still remains unsolved, especially for the high and precise demands in large-scale photonic integrated circuits (PICs). In this work, we develop the idea of manipulating the mode index of the partial length of a microring cavity by deposition or etching of material. By carefully adjusting the core waveguide mode's effective index of partial length, we discover a simple relation between the relative resonance frequency shift ($\delta f /f$) and the partial length ($L'$), as discussed in section 3. This relation of relative resonance frequency shift and partial length mode index manipulation provides an inherent advantage in the scalability of its method in a single fabrication step. We discuss our post-processing technique for tuning the resonance of a microring cavity and its effectiveness in minimizing the distribution of frequencies using analytical studies and numerical experiments. We use resonator and cavity terms interchangeably throughout this text.
\\

\noindent
A microring resonator is formed by shaping a waveguide into a loop or ring structure. This allows light to circulate within the loop through total internal reflection and interference at specific resonances. The resonance frequencies of the resonator are given by,
\begin{equation} \label{eq:1}
f_{m}=\frac{mc}{Ln_{eff}}
\end{equation}

\begin{itemize}
    \item \( f_{m} \) is the resonant frequency,
    \item \( m \) is the mode number,
    \item \( c \) is the speed of light in vacuum,
    \item \( L \) is the optical path length,
    \item $n_{eff}=n_{eff}(\omega)$ is frequency-dependent effective index.
\end{itemize}

\noindent
This equation indicates that the resonant frequency \( f_{m} \) is determined by the speed of light in the material, the effective refractive index \( n_{eff} \), the physical length of the optical path \(L \), and the mode number \( m \). We assume that the fabrication process errors affecting the distribution of resonances of cavities are small such that m is equal for a given resonance of all cavities in a large set of cavities.

\section{Resonance shift due to fabrication errors in $L$ and $n_{eff}$}

\begin{equation} \label{main_1}
f = \frac{mc}{L n_{eff}}
\end{equation}

\noindent
If there is a change in the length \( L \) by a small amount \( \delta L \), where \( \delta L \) can be negative or positive,  the new frequency \( f' \) is given by:

\begin{equation}
f' = \frac{mc}{(L + \delta L) n_{eff}}
\end{equation}

\noindent
For a small change in frequency we can assume, $n_{eff} (f) \sim n_{eff} (f')$, and the relative change in frequency \( \delta f /f\) is given by,

\begin{equation}
\frac{\delta f}{f} = \frac{f' - f}{f} = \frac{-\delta L}{L + \delta L}\sim \frac{-\delta L}{L}
\end{equation}


\noindent
Similarly, if there is a change in the effective index \( n_{eff} \) by a small amount \( \delta n_{eff} \), the new frequency \( f' \) and relative change in frequency, $\delta f /f$ are given by:

\begin{equation}
f' = \frac{mc}{L (n_{eff} + \delta n_{eff})}
\end{equation}

\begin{equation}
\frac{\delta f}{f} = \frac{f' - f}{f} = \frac{-\delta n_{eff}}{n_{eff} + \delta n_{eff}} \sim \frac{-\delta n_{eff}}{n_{eff}}
\end{equation}


\section{Partial length mode index $n_{eff}$ manipulation}
The effective mode index can be perturbed or manipulated by depositing a thin dielectric material or cladding on top of the cavity and etching of core waveguide (or cavity) or cladding material (assuming cladding thickness on top of the cavity is small). This method can be implemented to tune the cavity's resonance but doesn't provide the selective tuning advantage in a single fabrication step. Besides, by depositing a thin dielectric material or cladding film in a controlled manner on the partial length of the cavity, leaving the remaining cavity's length unperturbed, we can change the total optical path length of the mode in a controlled manner. By taking advantage of the lithography and dielectric thin film growth or deposition techniques, the dimensions of this layer can be formed on top of the cavity selectively to ultra-fine tune the total optical path length. Thus, it allows ultra-fine tuning of the cavity's resonance selectively. 
\\

\begin{figure}[ht!]
\centering\includegraphics[width=15.5cm]{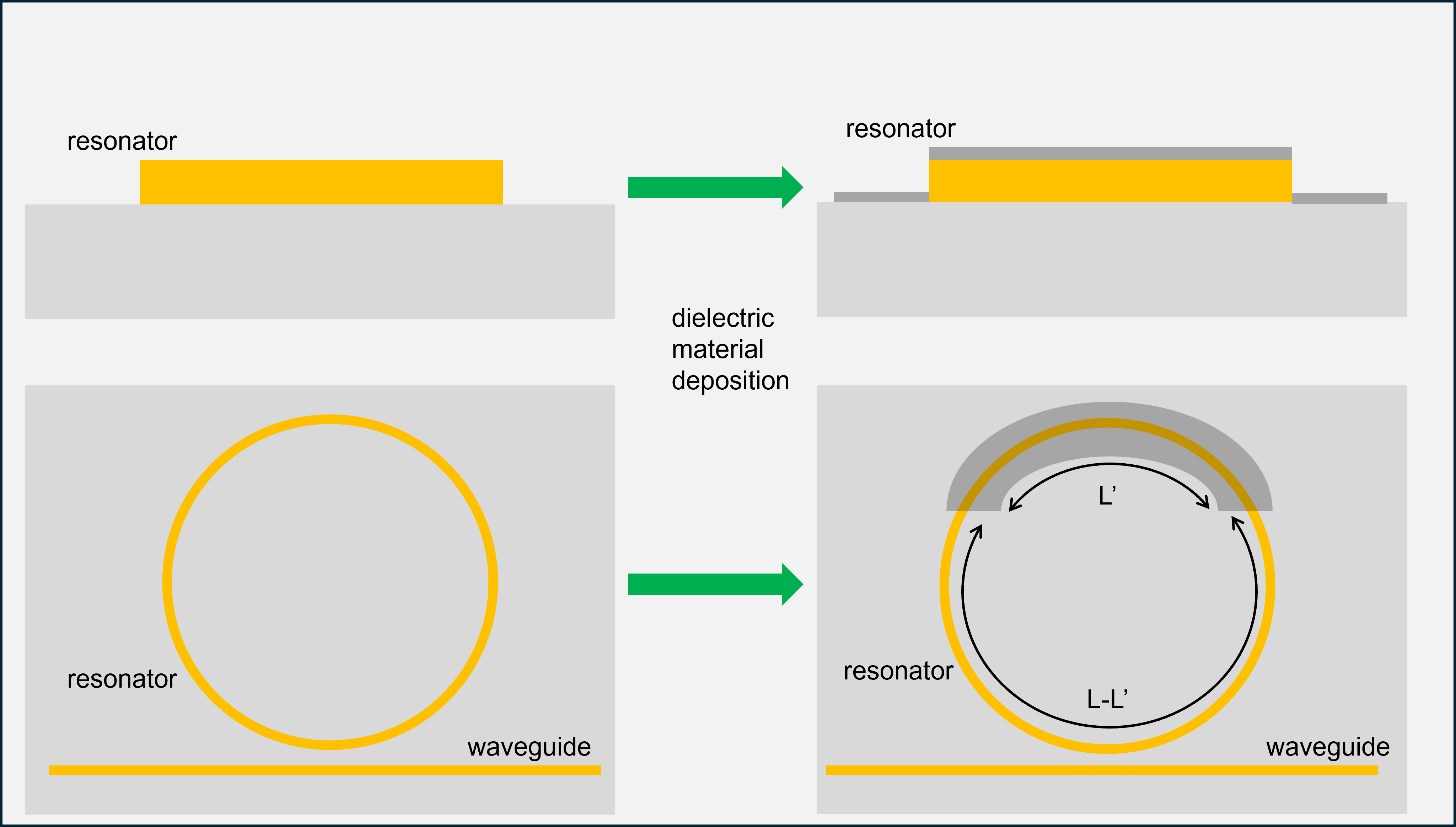}
\caption{Adding a dielectric layer of selective dimensions on top of the microring cavity to modify the total optical path of a mode.}
\label{cavity_tuning}
\end{figure}

\noindent 
We divide the total length L of the microring cavity into two segments $L'$ and $L-L'$, where $L-L'$ segment is unperturbed and has effective mode index $n_{eff}$ and $L'$ segment's effective mode index is changed from $n_{eff}$ to $n' _{eff}$ as shown in Fig. \ref{cavity_tuning}. For a small change in $n_{eff}$, i.e., $\delta n = n' _{eff} - n_{eff}$, m remains the same for a given resonance. The resonant condition for new arrangement is given by,
\begin{equation} \label{eq:2}
f'=\frac{mc}{L'n'_{eff} + (L-L')n_{eff}}
\end{equation}

\begin{figure}[ht!]
\centering\includegraphics[width=15.5cm]{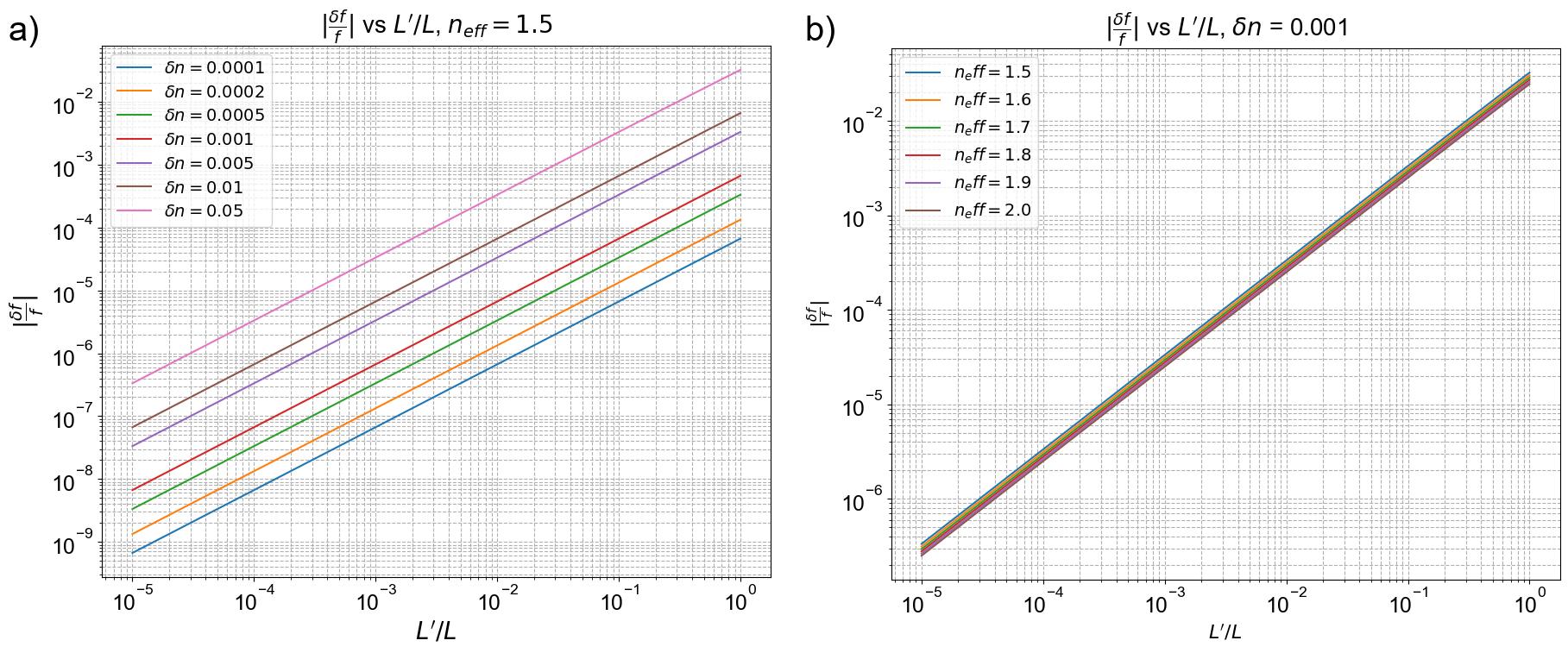}
\caption{ Plot of absolute relative change in frequency, |$\delta f/f$| with varied $L'/L$ ratio and a) $\delta n$ with $n_{eff}$ = 1.5 and b) $n_{eff}$ with $ \delta n$ = 0.001.}
\label{plot_1_2}
\end{figure}

\noindent
where $f'$ is the new resonance of the tuned cavity with $L'$ partial length of cavity's effective mode index $n_{eff}$ is changed to $n'_{eff}$ and $L-L'$ is the length of the cavity with unchanged effective index $n_{eff}$ and L is the total length of the cavity which is also unchanged. For the total length of cavity L and the effective index of mode  $n_{eff}$, the resonance frequency is given by eq. \ref{eq:1}, $f = \frac{mc}{Ln_{eff}}$. The relative change in resonance frequency is given by,

\begin{equation} \label{eq:5}
\frac{\delta f}{f} = \frac{f' - f}{f} =\frac{1}{1 + \frac{L'\delta n}{Ln_{eff}}} - 1
\end{equation}

\noindent
where $\delta n = n'_{eff} - n_{eff}$. When $\delta n$ is negative, the new frequency is higher than the original frequency, i.e., resonance is blueshift, and when $\delta n$ is positive, the new frequency is lower than the original frequency, i.e., resonance is redshift. For small values of $\delta n$, the eq. \ref{eq:5} can be simplified to, $\frac{\delta f}{f}\sim - \frac{L'\delta n}{Ln_{eff}}$. Using eq. \ref{eq:5}, we can evaluate the relative shift in resonance frequency when we change the effective mode index in the partial length $L'$ of a cavity. Fig. \ref{plot_1_2} shows the absolute relative change in frequency due to deposition of dielectric material on length $L'$. The smaller the change in index $\delta n$ and length ratio $L'/L$, the smaller the shift in resonance is achievable and vice versa.

\section{Passive Tuning of cavity's resonance Selectively through Mode Index Engineering}
\begin{figure}[ht!]
\centering\includegraphics[width=15.5cm]{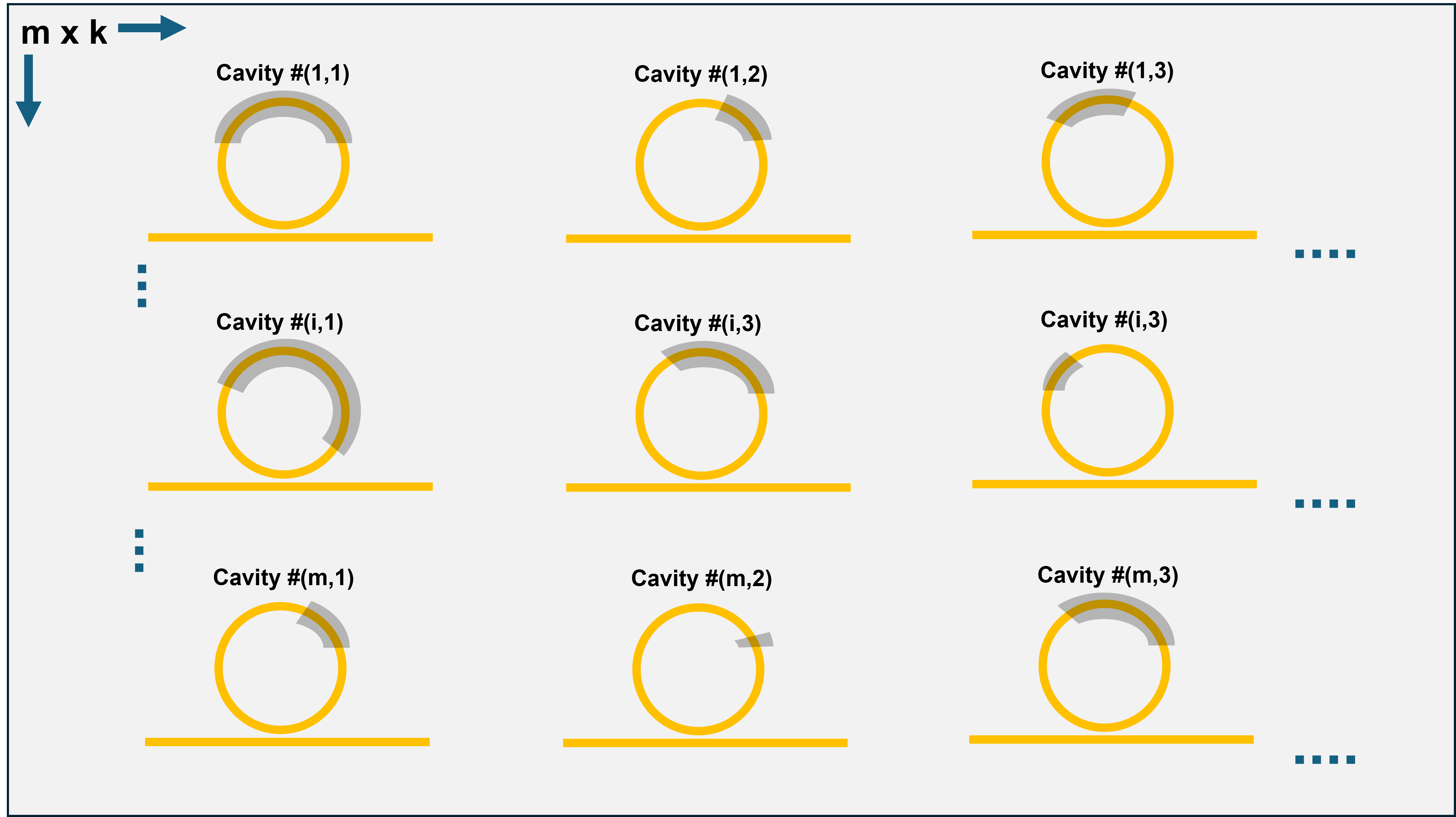}
\caption{Tuning the resonances of microring cavities selectively by individually selecting partial microring cavity length of each cavity. A set of one type cavities in an array of size m $\times$ k on a chip. The length ($L'$) is evaluated by the difference between the original frequency and the desired frequency, and detailed numerical analysis of $n_{eff}$, $\delta n$, and $\delta f(L')/f$ relation. The process can also be implemented on multiple chips with multiple cavities, thus enabling us to reduce frequency distribution from chip to chip in a large-scale manufacturing process.}
\label{cavities_tuning_resonance}
\end{figure}

We exploit the concept discussed above that the shift in resonance frequency of a microring cavity is achievable by introducing a change in the effective index of mode by thin film deposition of a dielectric material on top of the cavity (Fig. \ref{cavity_tuning}). We have the freedom to choose the length $L'$, i.e., the partial length of the microring cavity, to manipulate the effective index of mode selectively; we define this method of engineering the total optical path length as mode index engineering. Assuming that the cavities are open to air without any cladding or have minimal cladding on top such that if there is an introduction of dielectric material on top of the cavity, then it could achieve a change in the effective index of the mode significantly within a desired range of resonance shift, the tuning range and resolution can be understood from Fig. \ref{plot_1_2}. Fig. \ref{cavities_tuning_resonance} shows a schematic diagram of a set of cavities tuned selectively by the proposed idea of mode index manipulation of partial length. Due to the inherent scalability of this method, the tuning operation can be executed in a single fabrication process.

\subsection{Restricted red-shift or blue-shift tuning}
In principle, the effective mode index would increase if a dielectric material of index greater than the top cladding material (air, or dielectric) index is deposited on top of the cavity as discussed in Fig. \ref{cavity_tuning}; therefore, the cavity mode frequency would decrease. In minimizing the distribution of resonance frequencies by this method, we are restricted to red-shifting the resonances; therefore, the optimization in minimizing the frequency distribution of resonances is restricted towards the lowest frequency value in a set of cavities. On the other hand, etching or removing material from the resonator would decrease the effective mode index, and a blue shift in resonance frequency is achievable. Generally, the material etching is less precise than the industry-wide accepted deposition techniques, so we focus on the former in detail. Otherwise, the latter is also potentially feasible, and we can apply similar suitable design principles as well as numerical and experimental methods.

\clearpage
\subsection{Passive tuning of a silicon nitride resonator resonance}
\begin{figure}[ht!]
\centering\includegraphics[width=15.5cm]{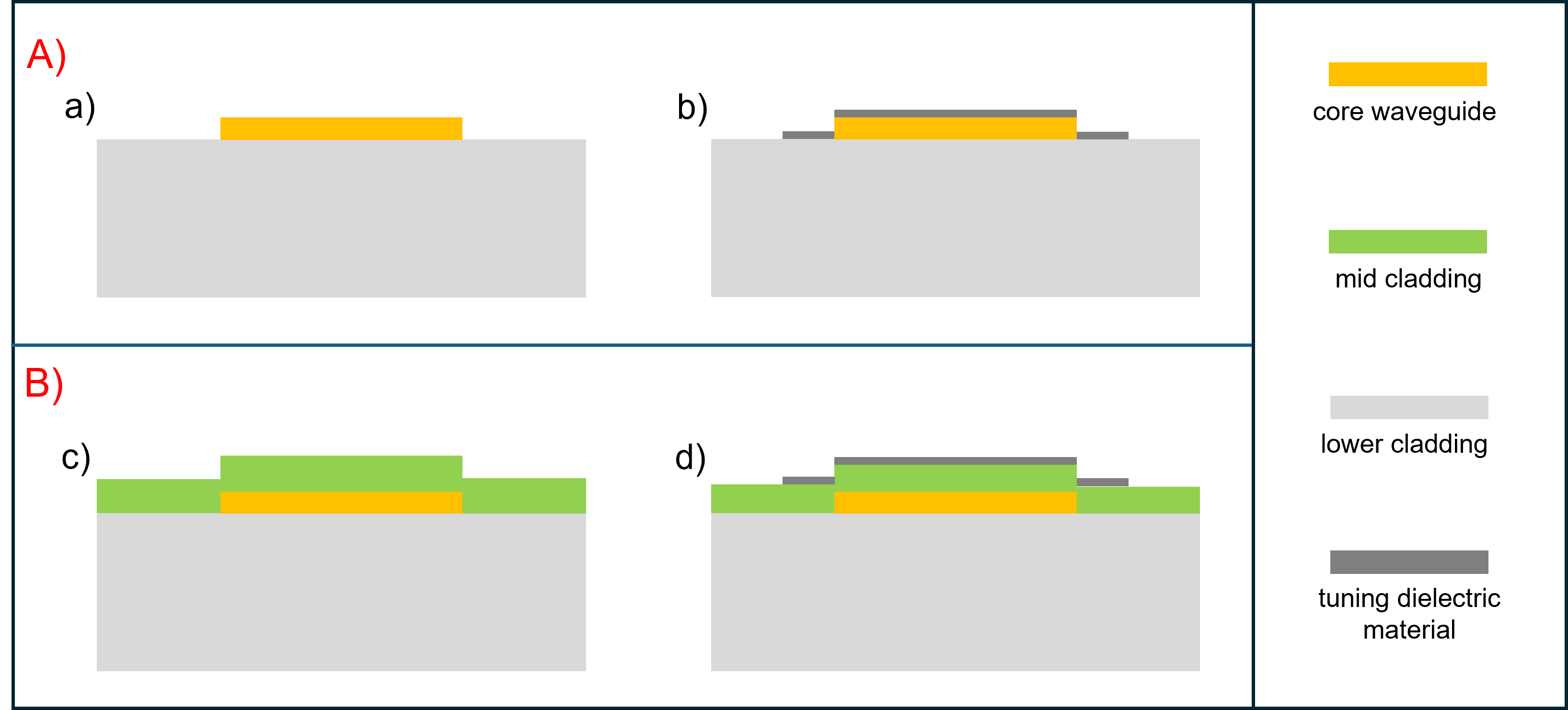}
\caption{The schematic diagram of the layer structure cross-sections of two configurations. First configuration (A): a) cavity core waveguide material $\text{Si}_{3}\text{N}_{4}$ (yellow color) on top of $\text{SiO}_{2}$ cladding (grey color). b) A dielectric material layer (dark grey color) is deposited on the partial length of the cavity to tune the cavity's resonance selectively (see Fig. \ref{cavity_tuning} and \ref{cavities_tuning_resonance}). Second configuration (B): c) cavity core waveguide material $\text{Si}_{3}\text{N}_{4}$ (yellow color) on top of $\text{SiO}_{2}$ cladding (grey color) with a cladding (green color) on top of $\text{Si}_{3}\text{N}_{4}$. d) A dielectric material layer (dark grey color) is deposited on a partial length of the cavity to tune the cavity's resonance selectively. In part c), the top cladding (green color) thickness should be small enough to enable the desired post-processing selective tuning of each cavity's resonance.}
\label{layer_struct_1}
\end{figure}

We discuss an example of silicon nitride ($Si_{3}N_{4}$) microring resonator and apply numerical analysis for red-shifting the resonance frequency. We use finite-difference time-domain (FDTD) analysis in our numerical experiments. In complementary metal-oxide-semiconductor (CMOS) technology, dielectric materials like silicon dioxide ($\text{SiO}_{2}$), aluminum oxide ($\text{Al}_{2}\text{O}_{3}$), and silicon oxynitride (SiON) are commonly used in photonic integrated circuits. We choose two configurations (A and B) for numerical demonstration purposes, as shown in Fig. \ref{layer_struct_1}. The first case (A) involves the original configuration of the microring resonator core waveguide being open to the air, while the second case (B) includes a cladding deposited on the microring resonator core waveguide layer structure. We estimate the change in the effective index of fundamental $TE_{0}$ mode due to the deposition of a thin film dielectric material and investigate the tunable range and resolution achievable by our method. Note that this method is applicable to any higher confined mode in a cavity.  We consider $\text{Si}_{3}\text{N}_{4}$ as core waveguide material and $\text{SiO}_{2}$ and $\text{Al}_{2}\text{O}_{3}$ as cladding materials. We assume the refractive index of $\text{Si}_{3}\text{N}_{4}$ is 2, $\text{Al}_{2}\text{O}_{3}$ is 1.76, and $\text{SiO}_{2}$ is 1.457 at 772.5 nm. Also, note that the material and geometry dispersion are included in the estimation of effective mode index $n_{eff}$. Therefore, the method applies to dispersive materials and accounts for material and geometric dispersions.

\begin{figure}[ht!]
\centering\includegraphics[width=15.5cm]{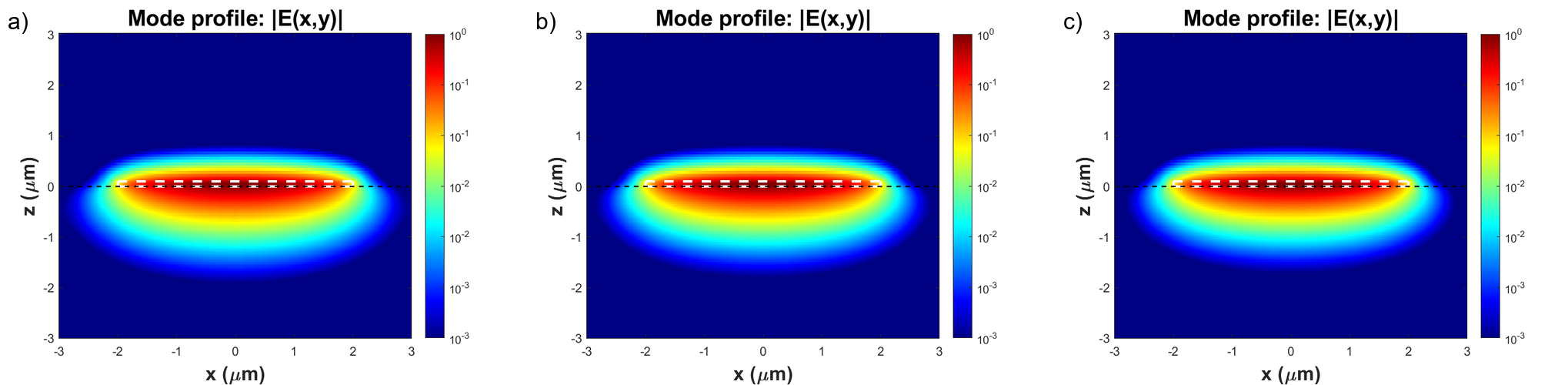}
\caption{Mode profile (|E(x,y)|) of $TE_{0}$ mode confined in 100 nm thick 4000 nm wide $\text{Si}_{3}\text{N}_{4}$ core waveguide for three cases: a) on $\text{SiO}_{2}$ cladding (see Fig. \ref{layer_struct_1}(a)), b) on $\text{SiO}_{2}$ cladding with $\text{SiO}_{2}$ thin film material of thickness 10 nm deposited on top  (see Fig. \ref{layer_struct_1}(b)), and c) on $\text{SiO}_{2}$ cladding with $\text{Al}_{2}\text{O}_{3}$ thin film material of thickness 10 nm deposited on top (see Fig. \ref{layer_struct_1}(b)). White and black dotted lines show $\text{Si}_{3}\text{N}_{4}$ core waveguide and lower $\text{SiO}_{2}$ cladding interface with $Si_{3}N_{4}$, respectively.}
\label{mode_profiles}
\end{figure}

\begin{figure}[ht!]
\centering\includegraphics[width=15.5cm]{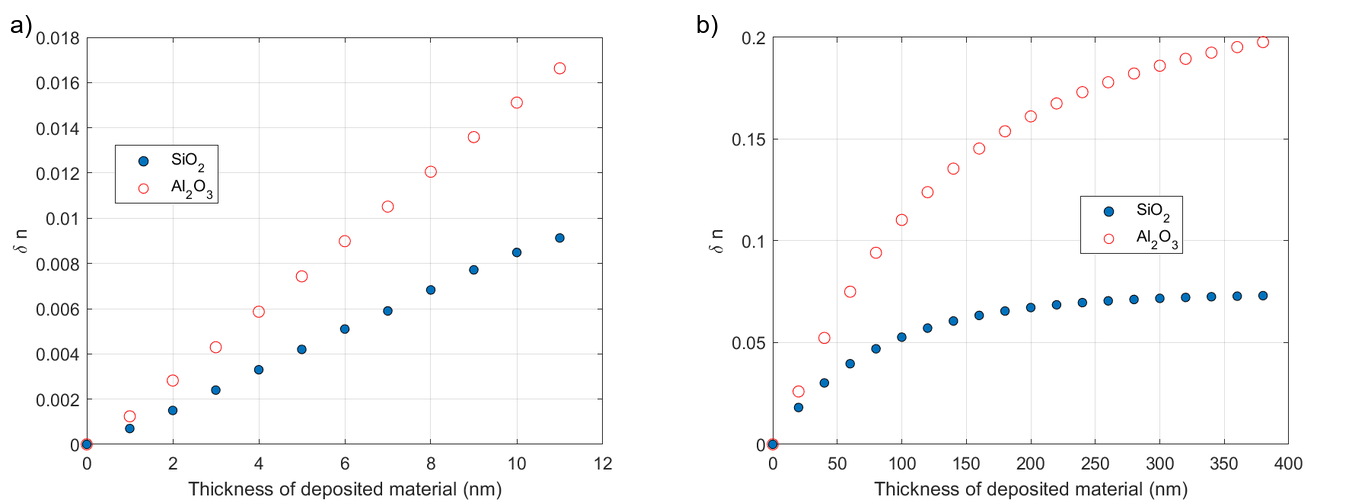}
\caption{The plot of change in effective mode index ($\delta n$) of $TE_{0}$ mode confined in 100 nm thick 4 $\mu$m wide $\text{Si}_{3}\text{N}_{4}$ core waveguide with varied thickness of deposited dielectric materials $\text{SiO}_{2}$ and $\text{Al}_{2}\text{O}_{3}$ (configuration A, see Fig. \ref{layer_struct_1}(b)): a) 0-11 nm thickness range on x-axis and b) 0-380 nm thickness range on x-axis.}
\label{effective_index_small_large_thickness}
\end{figure}

\begin{figure}[ht!]
\centering\includegraphics[width=15.5cm]{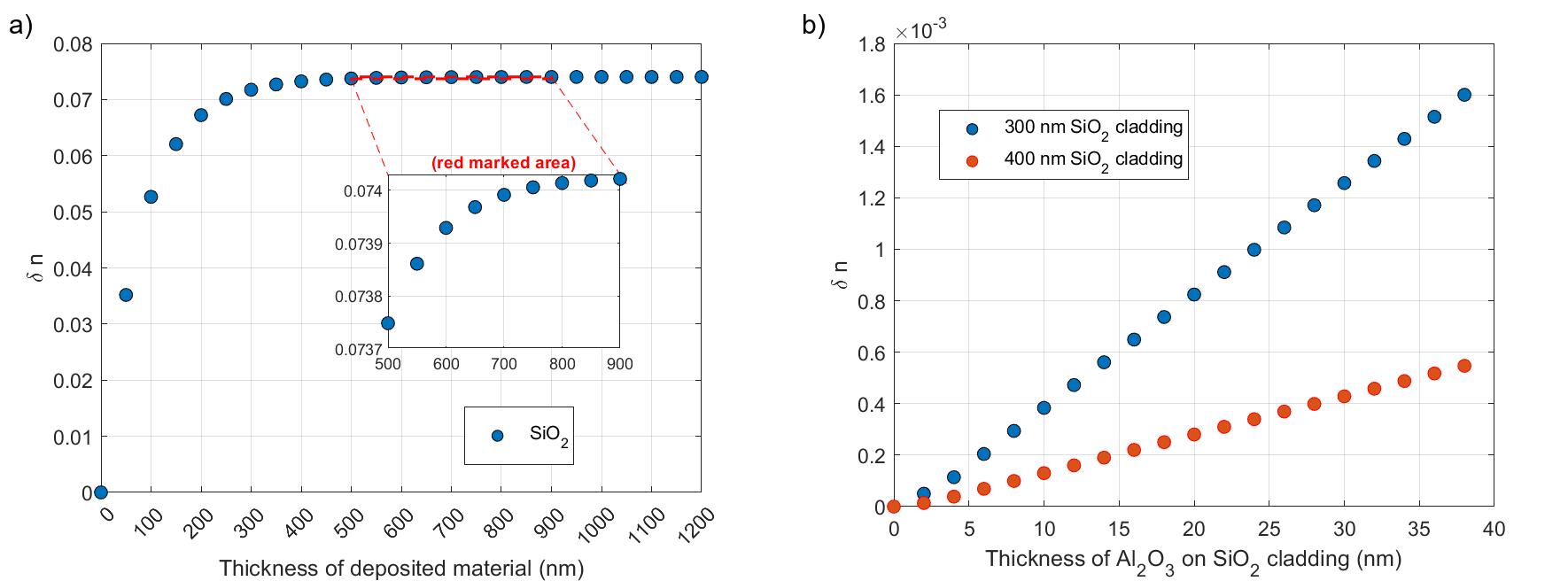}
\caption{Plot of change in effective mode index ($\delta n$) of $TE_{0}$ mode confined in 100 nm thick 4 $\mu$m wide $\text{Si}_{3}\text{N}_{4}$ core waveguide: a) due to deposition of $\text{SiO}_{2}$ dielectric material on top of $Si_{3}N_{4}$, $\text{Si}_{3}\text{N}_{4}$ sits on top of $\text{SiO}_{2}$ cladding (configuration A, see Fig. \ref{layer_struct_1}(b)). The inset shows a small range of data (red marked area); the change in $\delta n$ is achievable as small as $\sim 0.0001$  for even large $\sim$ 50 nm thick material deposition if configuration B (see Fig. \ref{layer_struct_1} (c) and (d)) is implemented, b) due to deposition of $\text{Al}_{2}\text{O}_{3}$ on top of 300 nm (blue dotted scatter points) and 400 nm (red dotted scatter points) thick $\text{SiO}_{2}$ cladding on $Si_{3}N_{4}$, $\text{Si}_{3}\text{N}_{4}$ sits on top of $\text{SiO}_{2}$ cladding (configuration B, see Fig. \ref{layer_struct_1} (c) and (d)). In this case, $\delta n \sim$ 0.0001 is attainable for about 5-10 nm thick $\text{Al}_{2}\text{O}_{3}$. The thickness of $\text{SiO}_{2}$ cladding can be optimized (see Fig. \ref{layer_struct_1}(c)) based on the desired $\delta n$ in the tuning operation.}
\label{array_4a}
\end{figure}

\noindent
In the numerical experiment, we assume a 100 nm thick 4000 nm wide $\text{Si}_{3}\text{N}_{4}$ core waveguide on top of $\text{SiO}_{2}$ cladding (lower cladding) for the microring resonator. Fig. \ref{mode_profiles}(a) shows the mode profile of fundamental TE (TE$_{0}$) mode confined in the core waveguide, Fig. \ref{mode_profiles}(b) shows the mode profile of TE$_{0}$ mode confined in $\text{Si}_{3}\text{N}_{4}$ core waveguide with 10 nm $\text{SiO}_{2}$ deposited on top and Fig. \ref{mode_profiles}(c) shows the mode profile of TE$_{0}$ mode confined in $\text{Si}_{3}\text{N}_{4}$ core waveguide with 10 nm $\text{Al}_{2}\text{O}_{3}$ deposited on top. Since the refractive index of $\text{Al}_{2}\text{O}_{3}$ is larger than $\text{SiO}_{2}$, the change in the effective index of mode is larger for a given thickness, as shown in Fig. \ref{effective_index_small_large_thickness}. Even for a small thickness $\sim$ 1 nm of deposited material (see Fig. \ref{effective_index_small_large_thickness} (a)), the change in the effective index, $\delta n$ is $\sim$ 0.001. In configuration B (see Fig. \ref{layer_struct_1}(c)), a certain thickness of cladding is on top of the core waveguide already, and the original resonance frequency is defined for this configuration B (Fig. \ref{layer_struct_1}(c)). The tuning step is performed by depositing a dielectric material (see Fig. \ref{layer_struct_1}(d)). As illustrated in Fig. \ref{effective_index_small_large_thickness} and \ref{array_4a}(a), the curve saturates for large thickness, and $\delta n$ can be an order of $\sim0.0001$ for the configuration B, which is suitable for ultra-fine tuning purposes. Fig. \ref{array_4a}(b) shows the change in the effective index of mode due to $\text{Al}_{2}\text{O}_{3}$ material deposition on a $\text{SiO}_{2}$ cladding of thickness 300 nm or 400 nm on top of the core waveguide. The change in $\delta n$ due to $\text{Al}_{2}\text{O}_{3}$ is small for 400 nm $\text{SiO}_{2}$ cladding compared to the 300 nm case, which is obvious due to the confinement of mode in the core waveguide and decrease in mode overlap. We can learn from this analysis and include the role of mid cladding thickness to optimize $\delta n$ for our specific application and needs.
\\

\noindent
The tuning range and resolution have a trade-off factor with each other and fabrication capabilities. For instance, the change in the effective index is $\sim$ 0.001 for the 1 nm layer deposited as evaluated in Fig. \ref{effective_index_small_large_thickness}(a). Let's assume that the length of the resonator is 1000 $\mu$m, and in the post-processing method, the photo-lithography dimensions of 250 nm ($L'$) x 10000 nm ($W'$) are feasible. So, we take $L' =  250$ nm for our case. Therefore, the smallest $L'/L$ feasible ratio is $2.5 \times 10^{-4}$, corresponds to tuning resolution, $|\frac{\delta f}{f}|$ is $\sim 1 \times 10^{-7}$ (see Fig. \ref{plot_1_2}). Let's take largest L'/L feasible ratio $\sim 0.25$, so the tuning range is $|\frac{\delta f}{f}|$ is $\sim 2 \times 10^{-4}$. Therefore, the larger the length of the resonator L and the smallest possible $L'$ in photolithography dimensions, the better the resolution and larger the tuning range are achievable. Since we have emphasized the deposition of dielectric material to increase the effective index of mode, the tuning of resonance of cavities is possible towards red-shift, i.e., towards the lower end of frequency distribution. There are trade-offs between the thickness, refractive index of the deposited layer, tuning resolution, and range. This process needs appropriate optimization based on the choice of materials, cavity dimensions, desired Q-factor, distribution of resonances, and available fabrication capabilities. In principle, achieving even further ultra-fine tuning and a wide tuning range is possible.
\\

\noindent
The second configuration (B) is described in Fig. \ref{layer_struct_1}(c), so the original frequency of the cavity's resonance is assumed for this configuration, and the desired frequency is set to this value for their application in PICs. The mode is confined in the core waveguide and the mode overlap with the top cladding is small. Adding another thin dielectric material on top of this cladding would produce a very small shift in the effective index of mode compared to the cavity without cladding on top. Fig. \ref{effective_index_small_large_thickness}(b) and \ref{array_4a}(a) shows the curve for change in the effective mode index with the cladding thickness deposited on top of 100 thick 4000 nm wide $\text{Si}_{3}\text{N}_{4}$ waveguide. The curve reaches saturation after a certain thickness. Before the saturation, the resonance is less sensitive. This region near saturation allows the ultra-fine tune of the cavity's resonance by choosing $\delta n \sim 0.0001$ or even smaller value appropriately. 

\subsection{Minimization of resonances frequencies distribution}
\begin{figure}[ht!]
\centering\includegraphics[width=15.5cm]{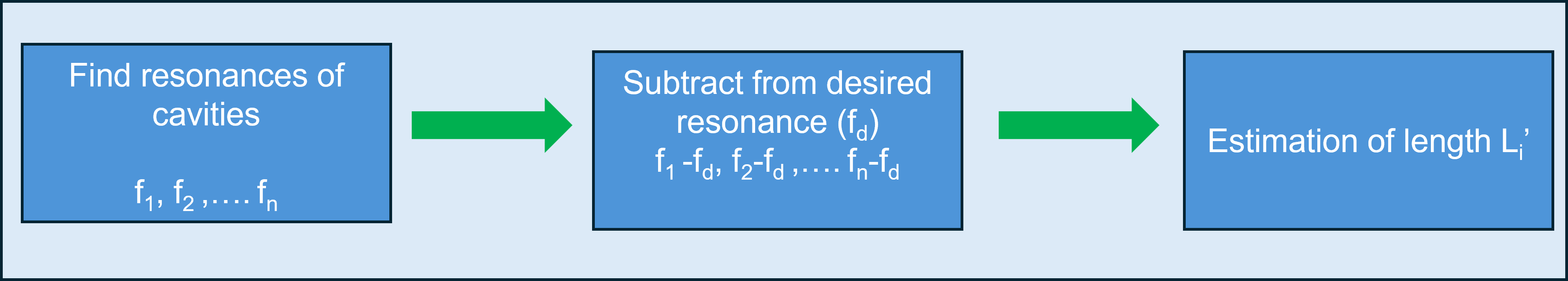}
\caption{Flow chart of steps to estimate $L'$.}
\label{algo}
\end{figure}
\noindent
In a large set of cavities, the estimation of partial length segments $L'_{i}$ for each cavity is estimated by evaluating $f_{i} - f_{d}$, (where $f_{i}$ is original or pre-tuned resonance frequency and $f_{d}$ is desired frequency) and using the analytical studies discussed in previous sections. The dimensions of the desired region for thin film dielectric material deposition can be estimated by examining $\delta n$. An exposure pattern is required for photo-lithography exposure on the photoresist. The desired dimensions and alignment at cavities can be achieved through computational capabilities in the lithography process. Fig. \ref{algo} shows the flow chart of steps for performing the optimization process in minimizing the distribution of frequencies. Note that since the deposition of the material increases the mode index, the resonances are red-shifted, so the desired $f_{d}$ is smaller than or equal to the smallest frequency in $f_{i}$ values. In principle, it is possible to implement this process in two or multiple steps to achieve the distribution as desired. In addition, this process can also be combined with a blue-shifting material etching process in multiple steps as desired.

\subsection{Losses: Fresnel reflection and Spatial Mode Overlap}

Two factors contribute to the losses in the applied method: Fresnel reflection at the transition from low index to high index and mode overlap, which happens at both transitions from low to high and high to low index regions. To estimate the Fresnel reflection coefficient (R) for different values of effective index $n_{eff}$, we can use the Fresnel equation for normal incidence,
\begin{figure}[ht!]
\centering\includegraphics[width=12cm]{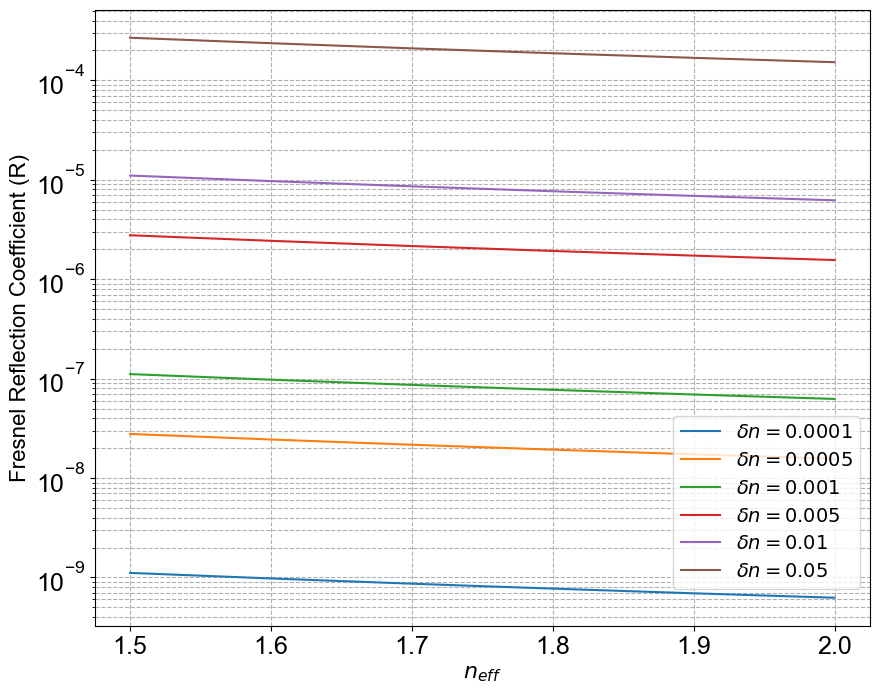}
\caption{Fresnel reflection coefficient (R) for varied $n_{eff}$ and $\delta n$.}
\label{fresnel}
\end{figure}

\begin{equation} \label{eq:fresnel}
R= \frac{(\delta n)^{2}}{(2n_{eff}+\delta n)^{2}}
\end{equation}

\noindent
The Fresnel reflection is significant for even a small value of $\delta n$. Spatial mode overlap is necessary at the transition of two regions, $L'$ and $L-L'$, for efficient light propagation at transitions. Both these losses cause a decrease in Q-factor and degrade the resonator's performance. Fig. \ref{fresnel} shows the evaluated R for varied $n_{eff}$ and $\delta n$. The spatial mode overlap can be numerically evaluated by the power coupling factor, $\kappa$, which evaluates the power coupled from one mode ($E_{1}$, $H_{1}$) to another mode ($E_{2}$, $H_{2}$) over the cross-section $dS$ integral area. The total area of the mode is spanned by area S.

\begin{equation}
\kappa = \frac{\left| \int_S \mathbf{E}_1^* \times \mathbf{H}_2 \, dS \right|}{\sqrt{\left( \int_S \mathbf{E}_1^* \times \mathbf{H}_1 \, dS \right) \left( \int_S \mathbf{E}_2^* \times \mathbf{H}_2 \, dS \right)}}
\end{equation}

\noindent
The ideal situation would be where $\kappa _{1}$ and $\kappa _{2}$ are equal to 1, where $\kappa _{1}$ is the power coupling factor from mode  transition $n_{eff}$ to $n'_{eff}$ region and $\kappa _{2}$ is the power coupling factor from mode  transition $n'_{eff}$ to $n_{eff}$ region. Although the change in the effective index between the two regions is small in our tuning method, it may be necessary to implement a tapering design depending on the Q-factor, $\delta n$, and thickness of deposited material. Generally, a large thickness of deposited material and $\delta n$ would produce a large mode mismatch. Therefore, employing techniques like tapered waveguides that improve spatial mode overlap to minimize optical power losses and prevent punishment on resonance Q-factor is important. In the next subsection, we discuss the application of adiabatic tapering in our design.

\subsection{Adiabatic tapering}
\begin{figure}[ht!]
\centering\includegraphics[width=15.5cm]{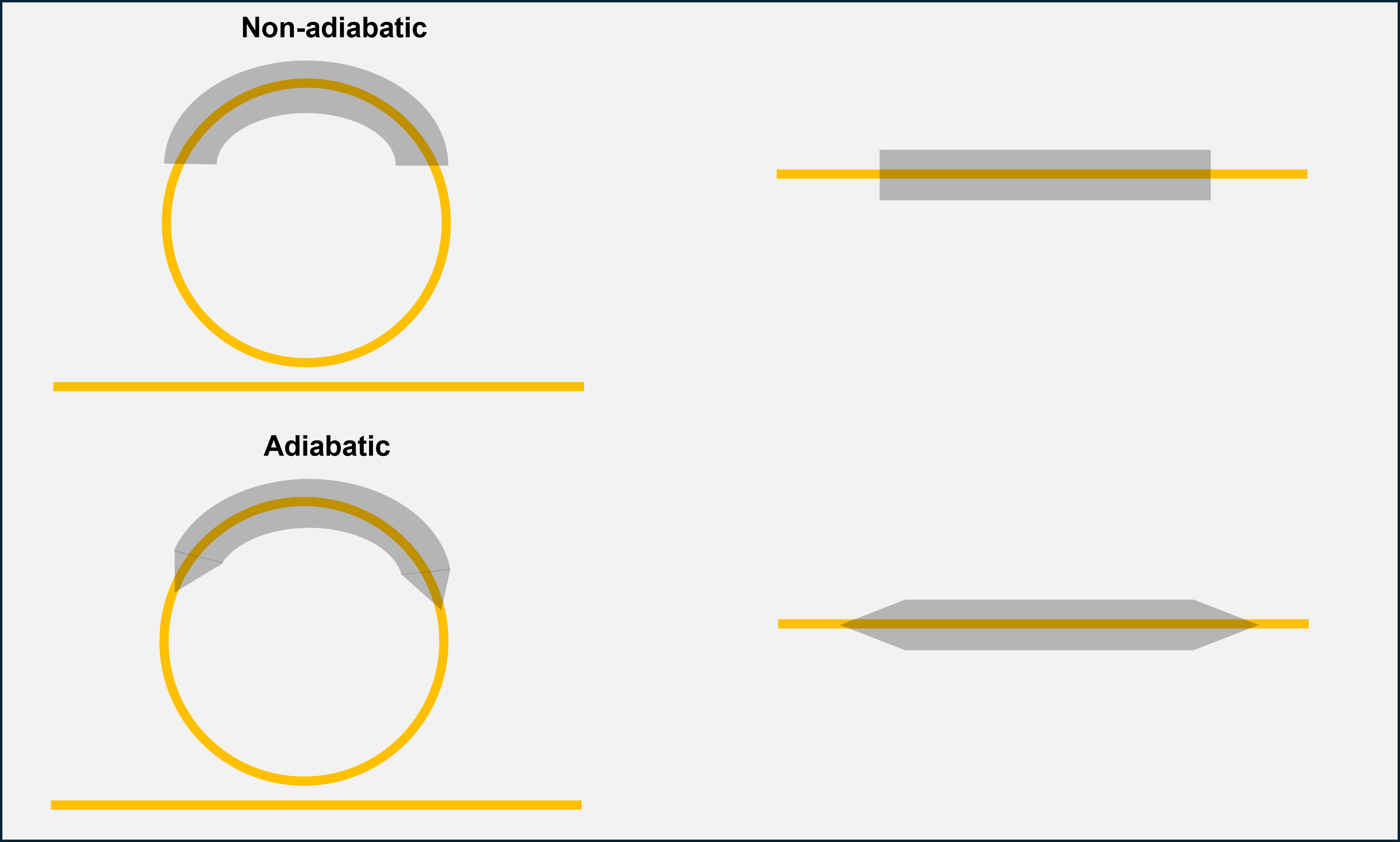}
\caption{Adiabatic region to tackle mode overlap and Fresnel reflection challenges. The grey color shows the deposited material on the top cavity material; in the non-adiabatic region ($L'$ segment), a rectangular shape is implemented, while the adiabatic design has two additional triangles at $L'$ segment ends.}
\label{adiabtic_tuning}
\end{figure}

Adiabatic tapering is essential in photonic waveguides to minimize Fresnel reflections and enhance mode overlap, which is critical for optimizing light transmission and efficiency. Adiabatic tapering gradually transitions the waveguide's cross-sectional area, reducing reflection losses and improving mode overlap. This enhances the overall performance and integration of photonic devices. Fig. \ref{adiabtic_tuning} shows the schematic diagram of the difference between non-adiabatic and adiabatic designs of deposited materials. The extra triangles, each of length $l$ at $L'$ segment ends, allow the adiabatic change in the effective index of the deposited material. Therefore, the effective mode indices of $L'$ and $L-L' -2l$ segments end transition into each other smoothly.

\section{Adiabatic design and resonance tuning}

In adiabatic tapering, the effective mode index \( n_{\text{eff}} \) changes gradually along the taper length \( x \) to minimize reflections and mode mismatch losses. While there isn't a single standard equation that universally defines \( n_{\text{eff}} \) as a function of length \( x \) during tapering, the evaluation generally depends on the specific geometry of the taper and the waveguide's material properties.
The evaluation of the effective index \( n_{\text{eff}} \) as a function of length in adiabatic tapering typically involves three key steps. First, the taper profile is designed, defining the shape (e.g., linear, exponential, or parabolic) that dictates how the waveguide width or height changes along its length, described by \( W(x) \), where \( W \) is the waveguide width at a position \( x \) along the length \( l \). Next, the effective index \( n_{\text{eff}}(x) \) is determined for each position \( x \) based on the local waveguide dimensions and material refractive indices, often using numerical simulations such as finite-difference time-domain (FDTD) or eigenmode solvers. Finally, the variation of \( n_{\text{eff}}(x) \) with respect to \( x \) is evaluated to understand how the mode evolves along the taper. Ensuring adiabaticity requires that the change in \( n_{\text{eff}}(x) \) is gradual enough to prevent mode coupling to higher-order or radiation modes.

\subsection{Linear Taper}
For a simple linear taper where the width changes linearly from \( W_1 \) to \( W_2 \) over a length \( l \), $W(x) = W_1 + \left(\frac{W_2 - W_1}{l}\right)x$. The effective index at position \( x \) can be expressed as,  $n_{\text{eff}}(x) \approx f(W(x))$, where \( f(W(x)) \) is the function that gives the effective index for a waveguide with width \( W(x) \). For the taper to be adiabatic, the rate of change of \( n_{\text{eff}} \) with respect to \( x \) should satisfy $\left|\frac{d n_{\text{eff}}(x)}{dx}\right| \ll \frac{n_{\text{eff}}(x)}{l}$. This ensures that the fundamental mode is preserved throughout the tapering process without significant coupling to other modes. Now, we discuss the effect of the additional tapering regions on the resonance frequency. The resonance condition is given by eq. \ref{main_1},
\[
f=\frac{mc}{Ln}
\]

\noindent
and for adiabatic tapered regions (a and b; a = ad.1 at one end and b = ad.2 at the second end), this equation becomes,
\begin{equation}
f'=\frac{mc}{L_{1}n_{eff} + L_{2}n'_{eff} + \int_{ad.1} n_{a}(x) dx + \int_{ad.2} n_{b}(x) dx}
\end{equation}

\noindent
where $L_{1}$ region has effective mode index $n_{eff}$, $L_{2}$ region has effective mode index $n'_{eff}$ and two regions (a and b) of length $l$ have effective mode index as a function of x, i.e., $n_{a}(x)$ and $n_{b}(x)$. Assuming the tapering region is linear of each adiabatic region of two ends (ad.1 and ad.2),

\begin{equation}
\int_{ad.1} n_{a} (x) dx = \int_{ad.2} n_{b} (x) dx= (n_{eff}+n'_{eff})l/2
\end{equation}
\begin{equation}
\frac{f'-f}{f} = \frac{\delta f}{f} =\left( \frac{L n_{eff}}{L_{1}n_{eff} + L_{2}n'_{eff} + (n_{eff}+n'_{eff})l}\right)  - 1  
\end{equation}

\noindent
where L is total length of cavity is, $L = L_{1} +L_{2} + 2l$. 

\begin{figure}[ht!]
\centering\includegraphics[width=15.5cm]{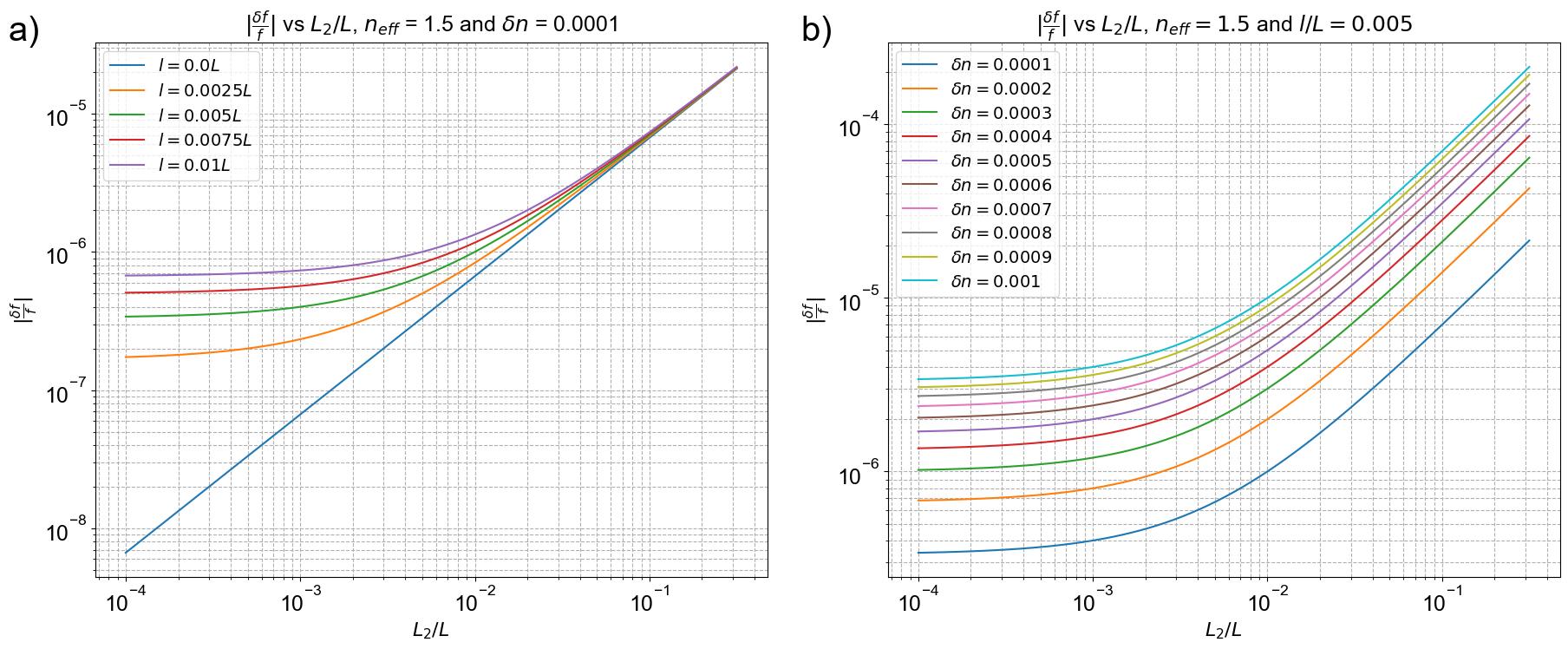}
\caption{Absolute relative change in frequency, |$\delta f/f$| with varied $L_2/L$ and a) $l/L$ with $\delta n$ = 0.0001 and b) $\delta n$ with $l = 0.005L$.}
\label{plot_4_5}
\end{figure}

\noindent
For $l = 0$, the equation reduces to 

\begin{equation}
\frac{f-f'}{f} = \frac{\delta f}{f} = \left( \frac{L n_{eff}}{L_{1}n_{eff} + L_{2}n'_{eff}}\right) - 1   
\end{equation}

\noindent
which is consistent with the main equation, $L = L_{1} + L_{2}$. Fig. \ref{plot_4_5} shows the plot of relative change in frequency at different ratios of $l/L$, $L_{2}/L$ and values of $n_{eff}$ and $\delta n$. The adiabatic region limits the tuning resolution and constrains the method's performance. Therefore, it is advised to optimize several factors or parameters, including materials choice, $\delta n$, $n_{eff}$, $l/L$, $L_{2}/L$, Q-factor, tuning range, tuning resolution and losses.

\section{Effect of cladding on tuned cavities}
Photonic integrated circuits require a cladding material on top to prevent any outside interaction with the photonic circuit elements and electromagnetic modes that could cause damage. Cladding on top of photonic-integrated devices reduces signal loss and crosstalk. It also offers protection against environmental damage, ensuring device reliability and durability. Therefore, it is essential to investigate the effect of thick cladding on top of a tuned microring cavity.

\begin{figure}[ht!]
\centering\includegraphics[width=15.5cm]{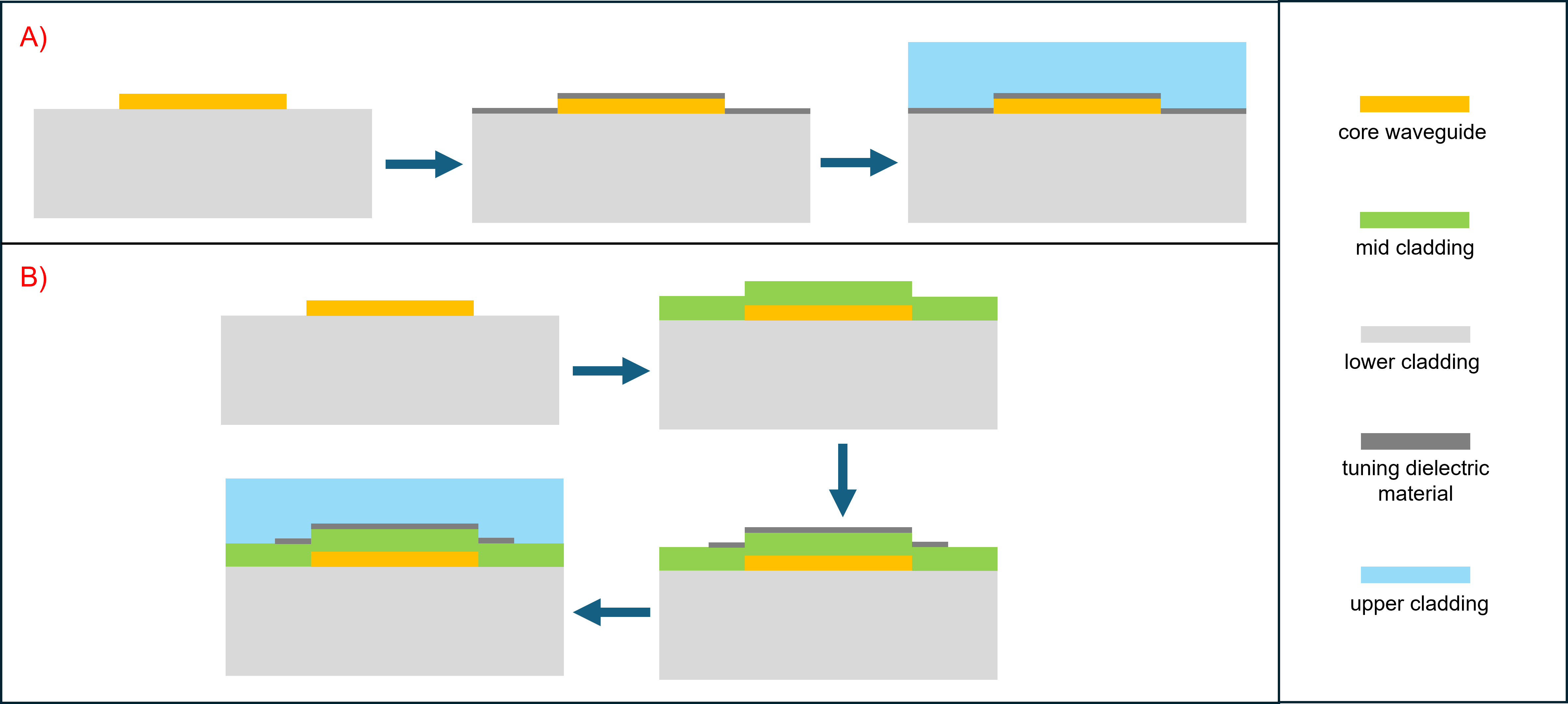}
\caption{The schematic diagram of the layer structure cross-sections of two configurations. First configuration (A): cavity core waveguide material $\text{Si}_{3}\text{N}_{4}$ (yellow color) on top of $\text{SiO}_{2}$ cladding (grey color). A dielectric material layer (dark grey color) is deposited on the partial length of the cavity to tune the cavity's resonance selectively. Finally, an upper cladding is added (blue color). Second configuration (B): cavity core waveguide material $\text{Si}_{3}\text{N}_{4}$ (yellow color) on top of $\text{SiO}_{2}$ cladding (grey color) with a mid-cladding (green color) on top of $\text{Si}_{3}\text{N}_{4}$. A dielectric material layer (dark grey color) is deposited on a partial length of the cavity to tune the cavity's resonance selectively. Finally, an upper cladding is added (blue color).}
\label{Cladding_on_tuned_cavities}
\end{figure}

\noindent
The addition of cladding on top of the tuned cavity changes the effective mode index and resonance shift from $f'$ to $f_{cladding}$. So, the resonance frequency shifts in two steps, 

$$f \rightarrow f' \rightarrow f_{cladding} $$
\noindent where $f'$ is tuned cavity resonance.

$$f = \frac{mc}{Ln} \rightarrow f' = \frac{mc}{(L-L')n + L'(n+\delta n)} \rightarrow f_{cladding} $$

\noindent For any tuned cavity, $(L-L')n + L'(n+\delta n) = \text{constant} = K$. After adding a cladding on top i.e. upper cladding,

$$\text{effective mode index of L-L' segment changes}:\,\,\, n \rightarrow n'$$

$$\text{effective mode index of L' segment changes}:\,\,\,n + \delta n \rightarrow n_{p}$$

\noindent
For small values of $(n_{p}-\delta n -n)$, and $(n'-n)$, mode number $m$ remains the same. In Fig. \ref{Cladding_on_tuned_cavities}, two configurations are shown, A and B; in the case of A, $\delta n$ is large and limits high tuning resolution, but in B, $\delta n$, $\delta \alpha$ and $\delta \beta$ are small numbers, high-Q resonators are fine-tuned i.e. configuration B is more favorable for tuning procedure because of the small $\delta n$, $\delta \alpha$, and $\delta \beta$. Therefore, We limit the numerical experiment to configuration B only.  Now we evaluate, $\frac{\Delta f'_{cladd.}}{f'} = \frac{f_{cladding} - f'}{f'}$,

$$\frac{\Delta f'_{cladd.}}{f'} = \frac{(L-L')n + L'(n+\delta n)}{(L-L')n' + L'n_{p}} - 1 = \frac{K}{(L-L')n' + L'n_{p}} - 1$$

\noindent We introduce two new parameters $\delta \alpha$ and $\delta \beta$,
$$n_{p} = n + \delta n + \delta \alpha$$

$$n' =  n + \delta \beta$$

\noindent
where $\delta \alpha$ is the change in the effective mode index of the L' segment, and $\delta \beta$ is the change in the effective mode index of the L-L' segment after adding a cladding on the tuned cavity.

$$\frac{\Delta f'_{cladd.}}{f'} = \frac{(L-L')n + L'(n+\delta n)}{(L-L')(n+\delta \beta) + L'(n+\delta n +\delta \alpha)} - 1= \frac{K}{(L-L')(n+\delta \beta) + L'(n+\delta n +\delta \alpha)} - 1$$

$$\frac{\Delta f'_{cladd.}}{f'} = \frac{K}{K + (L-L')(\delta \beta) + L'(\delta \alpha)} - 1$$

\noindent
Ideally, we want $\frac{\Delta f'_{cladd.}}{f'}$ to be independent of L' so that adding an upper cladding doesn't introduce a worsening of frequency distribution, though the resonance will shift. Only two factors, $\delta \alpha$ and $\delta \beta $, contribute to the shift of resonance and worsening of frequency distribution due to the addition of cladding on top of tuned resonators ($K$ is constant for all tuned cavities).

The critical quantity is, $\Delta M$ which is defined as, $$\Delta M = \left(\frac{\Delta f'_{cladd.}}{f'}\right)_{L'=L}  - \left(\frac{\Delta f'_{cladd.}}{f'}\right)_{L'=0}$$ This evaluates the maximum change in frequency distribution due to addition of an upper cladding.

$$\Delta M =  \frac{1}{(1 +\frac{\delta \alpha}{n +\delta n})} - \frac{1}{(1+ \frac{\delta \beta}{n})}$$

$$\Delta M \sim \frac{ ( \delta \beta -\delta \alpha)}{n + (\delta \beta + \delta n +\delta \alpha)} \sim \frac{ ( \delta \beta -\delta \alpha)}{n}$$

Only the magnitude of $\Delta M$ is relevant. Ideally, we want this quantity to be below 1/Q or tuning resolution. These numbers depend on the cavity's material configuration and should be evaluated using numerical methods such as finite-difference time-domain (FDTD) analysis. For an example of a numerical experiment, we consider a 100 nm thick 4000 nm wide silicon nitride ($Si_{3}N_{4}$) core waveguide on silicon dioxide ($\text{SiO}_{2}$) cladding, which we call lower cladding. Another $\text{SiO}_{2}$ cladding is on top of $\text{Si}_{3}\text{N}_{4}$ material, which we call mid cladding, and $\text{Al}_{2}\text{O}_{3}$ layer is used for tuning step, and the top cladding is called as upper cladding. The refractive index of $Si_{3}N_{4}$, $\text{SiO}_{2}$ and $\text{Al}_{2}\text{O}_{3}$ are assumed as 2.000, 1.457, and 1.76 respectively at 772 nm. The simulation results are estimated at the operating wavelength, 772 nm, for fundamental quasi-transverse electric (TE$_{0}$) mode.

\begin{figure}[ht!]
\centering\includegraphics[width=15.5cm]{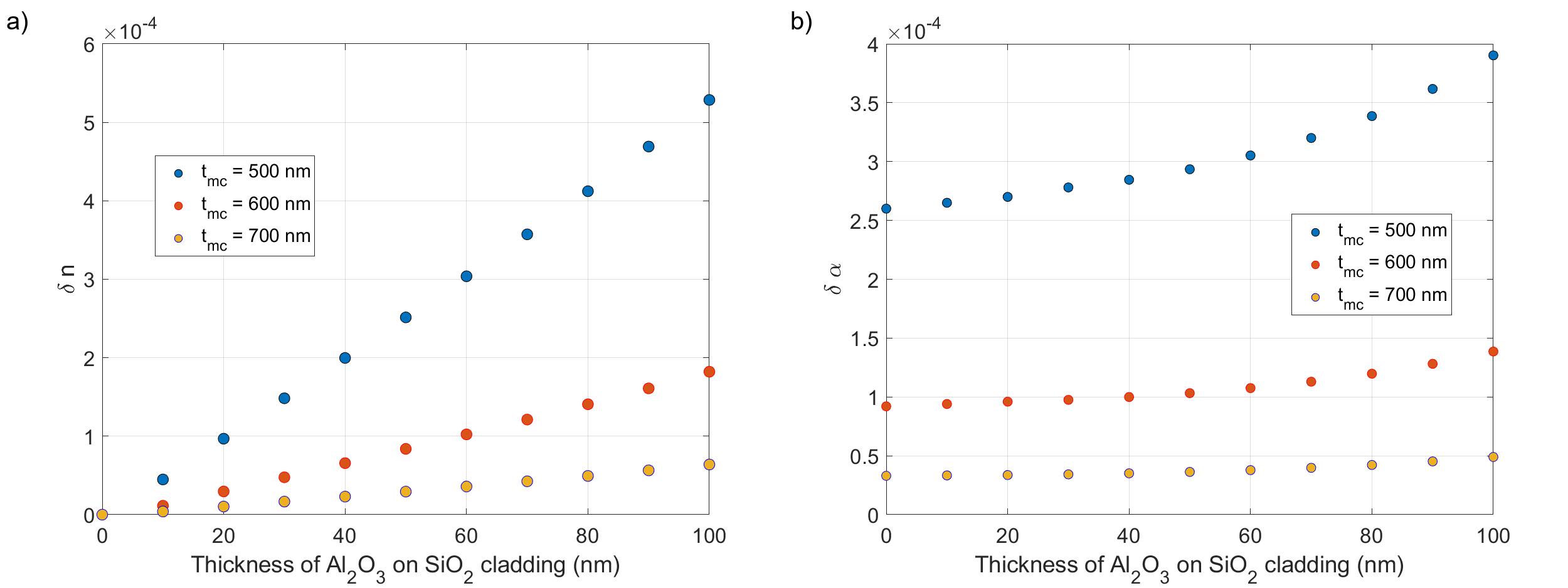}
\caption{a) Plot of $\delta n$ vs. thickness of $\text{Al}_{2}\text{O}_{3}$ layer on $\text{SiO}_{2}$ cladding without an upper cladding. $t_{mc}$ is the mid $\text{SiO}_{2}$ cladding thickness. b) Plot of $\delta \alpha$ vs. thickness of $\text{Al}_{2}\text{O}_{3}$ when an upper cladding $\text{SiO}_{2}$ is added on top. $\delta \alpha$ is evaluated by estimating the effective mode index (with upper cladding) and subtracting the effective mode index (without upper cladding). Note that, at zero thickness of $\text{Al}_{2}\text{O}_{3}$ i.e. $\delta n =0$, $\delta \beta$ = $\delta \alpha$. For both (a) and (b), three different thicknesses of mid $\text{SiO}_{2}$ cladding ($t_{mc}$) are considered, i.e., 500 nm, 600 nm, and 700 nm. For the larger value of $t_{mc}$, smaller values of $\delta n$ and $\delta \alpha$ are found which is obvious due to less mode overlap with $\text{Al}_{2}\text{O}_{3}$ and upper cladding as $t_{mc}$ increases. But the critical parameter is $\Delta M$, which is dependent on $\delta \alpha -\delta \beta$.}
\label{array_a}
\end{figure}

\begin{figure}[ht!]
\centering\includegraphics[width=15.5cm]{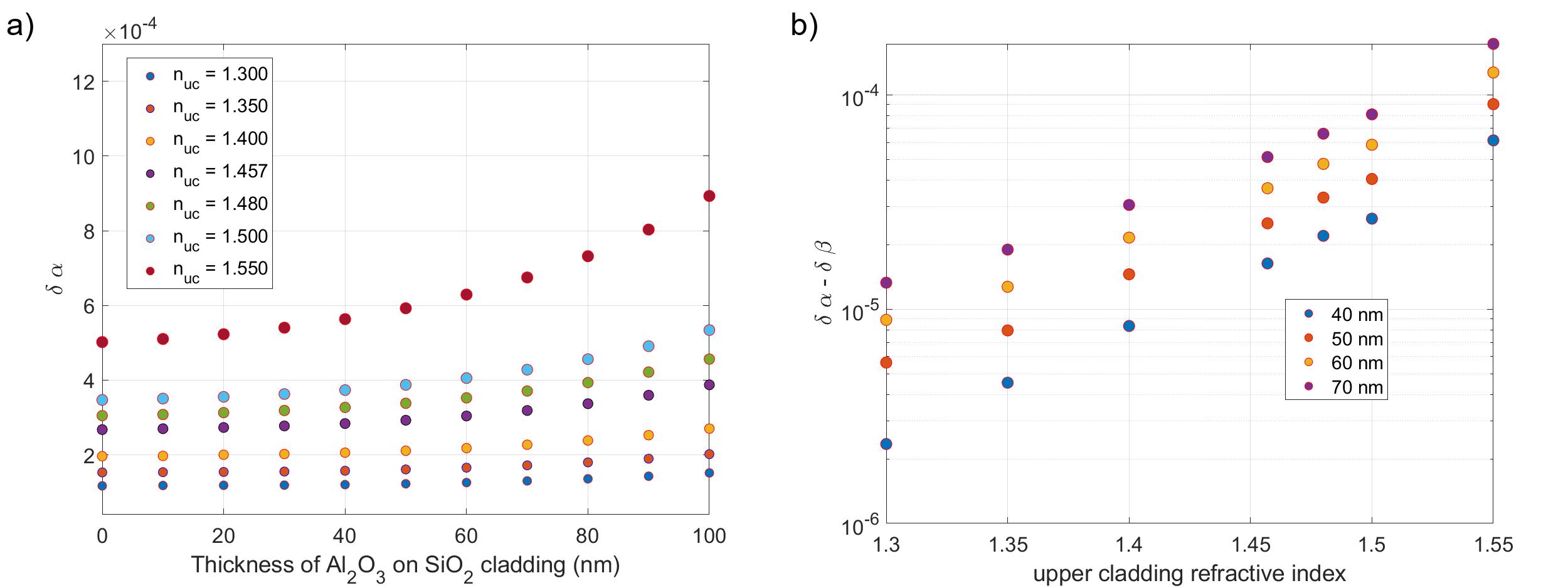}
\caption{Upper cladding refractive index ($n_{uc}$) dependent $\delta \alpha$ and $\delta \alpha - \delta \beta$ for $t_{mc}$ = 500 nm, a) Plot of $\delta \alpha$ vs. thickness of $\text{Al}_{2}\text{O}_{3}$ on 500 nm $\text{SiO}_{2}$ mid cladding and b) Plot of $\delta \alpha - \delta \beta$ vs. $n_{uc}$ for 40 nm, 50 nm, 60 nm, and 70 nm thick $\text{Al}_{2}\text{O}_{3}$ on 500 nm $\text{SiO}_{2}$ mid cladding; $\delta \beta$ is calculated at zero thickness of $\text{Al}_{2}\text{O}_{3}$ from (a).}
\label{array_b}
\end{figure}

\noindent
Fig. \ref{Cladding_on_tuned_cavities} shows two configurations, A and B. When the thickness of mid-cladding ($t_{mc}$) decreases, the mode overlaps with tuning dielectric and upper cladding material increases, and $\delta n$, $\delta \alpha$, and $\delta \beta$ increase; configuration B becomes A when the thickness of mid-cladding is zero. Fig. \ref{array_a}(a) and (b) shows the dependence of $\delta n$, $\delta \alpha$ on $t_{mc}$ and thickness of tuning dielectric material $\text{Al}_{2}\text{O}_{3}$, respectively. Note that values of $\delta \alpha$ (see Fig. \ref{array_a}(b)) are similar in order of $\delta n$ when an $\text{SiO}_{2}$ upper cladding is added. But when the $t_{mc}$ is large the variation of $\delta \alpha $ is small with thickness of $\text{Al}_{2}\text{O}_{3}$ or $\delta n$. This is an important observation; a relatively larger value of the mid-cladding and tuning dielectric material thicknesses smaller values are found for $\delta \alpha$ and $\delta \beta$ (see Fig. \ref{array_a}(b), \ref{array_b}(a) and (b)). Fig. \ref{array_b}(a) and (b) shows the $\delta \alpha $ and $\delta \alpha -\delta \beta$ dependence on the thickness of tuning dielectric material $\text{Al}_{2}\text{O}_{3}$ and upper cladding refractive index ($n_{uc}$) for $t_{mc}$ = 500 nm. Note that as the upper cladding refractive index decreases, $\delta \alpha - \delta \beta$ decreases significantly, favoring the mode index engineering method for tuning applications.
In the discussed example, $\Delta M$ is $\sim 10^{-6}$ for $\delta n \sim 2 \times 10^{-4}$ and $n_{uc}$ = 1.3 - 1.35. Therefore, the method discussed here attains high-resolution tuning and a wide tuning range for wafer-scale manufactured microring cavities with loaded-Q factor in the order of 1 million. In principle, even higher Q resonators can be tuned with appropriate optimizations of materials and use of low-index upper cladding materials.

\noindent
Now, we work out the $\Delta M$ quantity for adiabatic design. For adiabatically tapered regions (a and b; a = ad.1 at one end and b = ad.2 at the second end),

$$f'=\frac{mc}{(L-L'-2l)n + L'n' + \int_{ad.1} n_{a}(x) dx + \int_{ad.2} n_{b}(x) dx}$$

\noindent
where $L-L'-2l$ region has effective mode index $n$, $L'$ region has effective mode index $n'$ and two regions (a and b) of length $l$ have effective mode index as a function of x, i.e., $n_{a}(x)$ and $n_{b}(x)$. Assuming the tapering region is linear of each adiabatic region of two ends (ad.1 and ad.2),

$$\int_{ad.1} n_{a} (x) dx = \int_{ad.2} n_{b} (x) dx= (n+n')l/2$$

\noindent
and after the addition of cladding, $n_{a}(x)$ and $n_{b}(x)$ becomes $n'_{a}(x)$ and $n'_{b}(x)$ respectively, therefore,

$$\int_{ad.1} n'_{a} (x) dx = \int_{ad.2} n'_{b} (x) dx= (n'+n_{p})l/2$$

\noindent
Therefore,

$$\frac{\Delta f'_{cladd.}}{f'} = \frac{(L-L'-2l)n + L'(n+\delta n)+(2n+\delta n)l}{(L-L'-2l)n' + L'n_{p}+(n'+n_{p})l} - 1$$

\noindent The quantity $\Delta M$ is estimated as,

$$\Delta M = \left(\frac{\Delta f'_{cladd.}}{f'}\right)_{L'+2l = L}  - \left(\frac{\Delta f'_{cladd.}}{f'}\right)_{L'=0, l=0}$$

$$\Delta M \sim \frac{\delta \beta- \delta \alpha}{n}$$

\noindent
which is similar to the $\Delta M$ for non-adiabatic design. This is because only the difference of $\delta \beta$ and $\delta \alpha$ contribute significantly to $\Delta M$, and $l$ is significantly smaller compared to $L$.

\section{Discussion and Conclusion}
\noindent
Note that the error in thin film growth with atomic layer deposition is typically 1\%; therefore, the tuning step of deposition of the thin film introduces an error in a relative frequency shift of only 1\%. This is understood from an error in $\delta n$ from the desired value. The well-known and established techniques for high-quality thin film dielectric mirrors achieve such a level of accuracy in thin film deposition. Future advancements in high-accuracy thin film growth would propel the advantages of the method discussed here. 
\\

\noindent
We have discussed the concept of mode index engineering and its application in tuning the resonance of a cavity. We performed an analytical study of the method and provided appropriate steps to effectively minimize the distribution of resonances in a set of cavities. We applied this method with numerical analysis on a silicon nitride microring resonator in two configurations without cladding and with a thin cladding on top of the core waveguide material. We discussed the occurrence of losses from mode overlap mismatch and Fresnel reflection and presented its solution by implementing an adiabatic design. 
\\

\noindent
We introduced new parameters and quantities to estimate the worsening of frequency distribution due to cladding deposition on top of tuned cavities. It is found that the mode index-engineering method for passive tuning applications allows a scalable single-step tuning process for high-Q microring resonators. In the discussed numerical experiment, resonators with loaded Q-factor up to $10^{6}$ can be tuned with resolution below 1/Q and tuning range up to $(10^{2} - 10^{3})/Q$ without any significant impact on the frequency distribution after the addition of low index upper cladding. Low index materials for upper cladding are favorable to minimize the $\Delta M$; this is evident due to the high index contrast between the mid-cladding and upper-cladding material. Therefore, it is essential to review low-index materials that are optically transparent, low-cost, and compatible with complementary metal-oxide-semiconductor (CMOS) technology. The high-quality growth of such low-index materials on top of materials used in the mid-cladding and tuning step is a crucial area for future research.
\\

\noindent
We envision that the idea discussed here will be extended in further research including experiments and tuning the photonic crystal cavity's resonances.




\section{Author Contributions}
M.K. conceived the idea, performed the numerical simulations, and wrote the manuscript. S.D. and Z.Y. contributed to the discussions and revised the manuscript. 


\section{Funding}
S.D. is supported by Herman F. Heep and Minnie Belle Heep Texas A\&M University Endowed Fund held/administered by the Texas A\&M Foundation. We want to thank the Robert A. Welch Foundation (grants A-1261 and A-1547), the DARPA PhENOM program, the Air Force Office of Scientific Research (Award No. FA9550-20-10366), and the National Science Foundation (Grant No. PHY-2013771). This material is also based upon work supported by the U.S. Department of Energy, Office of Science, Office of Biological and Environmental Research under Award Number DE-SC-0023103, DE-AC36-08GO28308. 

\section{Acknowledgment}
We express our gratitude to Prof. Marlan O. Scully for the insightful discussions during the course of this study.

\section{Disclosures}
The authors declare that they have no known competing financial interests or personal relationships that could have appeared to influence the work reported in this paper.

\section{Data Availability}
Data underlying the results presented in this paper are not publicly available at this time but can be obtained from the authors upon reasonable request.


\bibliography{sample}

\end{document}